\documentclass[aps,pre,twocolumn,groupedaddress]{revtex4-1}
\usepackage{amsmath}
\usepackage{graphicx}
\begin{document}
\title{Tracer dynamics in single-file system with absorbing boundary}
\author{Artem Ryabov}
\email[]{rjabov.a@gmail.com}
\author{Petr Chvosta}
\email[]{petr.chvosta@mff.cuni.cz}
\affiliation{Charles University in Prague, Faculty of Mathematics and Physics, Department of Macromolecular Physics,
V Hole{\v s}ovi{\v c}k{\' a}ch 2, 18000 Praha 8, Czech Republic}
\date{\today}
\begin{abstract}
The paper addresses the single-file diffusion in the presence of an absorbing boundary. The emphasis is on
an interplay between the hard-core interparticle interaction and the absorption process. The resulting dynamics exhibits several qualitatively
new features. First, starting with the exact probability density function for a given particle (a tracer), we study the long-time asymptotics of its moments.
Both the mean position and the mean square displacement are controlled by dynamical exponents which depend on the initial order of the particle in the file.
Secondly, conditioning on non-absorption, we study the distribution of long-living particles.
In the conditioned framework, the dynamical exponents are the same for all particles, however, a given particle possesses an effective diffusion coefficient
which depends on its initial order. After performing the thermodynamic limit, the conditioned dynamics of the tracer is subdiffusive, the generalized diffusion coefficient $D_{1/2}$
being different from that reported for the system without absorbing boundary.
\end{abstract}
\maketitle

\section{Introduction}
Consider a single particle diffusing in the semi-infinite one-dimensional channel. The particle escapes from the channel only if it hits the channel boundary situated at the origin. Assuming normal diffusion without any external drift, the mean particle position remains constant in time, its mean escape time is infinite, nevertheless the particle eventually escapes with the probability one. At a given time $t$, starting with an ensemble of all possible particle trajectories, it is interesting to restrict the attention on the subensemble of those paths which have not hit the boundary up to the time $t$. The subensemble is described by the {\em conditional\/} probability density functions (PDFs), the condition being the non-absorption up to the time $t$. The conditioned dynamics exhibits qualitatively different features comparing with the unconditioned one. For instance, the conditioned mean particle position is no longer constant, it growths as $t^{1/2}$. The longer one waits the further from the origin is the typical surviving trajectory. One can say that the conditioning implies an {\em effective\/} force which drags the particle away from the absorbing boundary.

In the present paper we shall investigate the system of $N$ hard-core interacting particles diffusing in the semi-infinite one-dimensional channel with the absorbing boundary at the origin.
We have three main objectives. First, the hard-core interaction implies an entropic interparticle repulsion and we analyze its effect on the dynamics of the individual particle (a {\em tracer\/}). Second, we are interested in the dynamics of the long-living particles, that is, in the tracer dynamics conditioned on non-absorption. Third, we shall compare the dynamics of the system of $N$ particles with that of the corresponding system in thermodynamic limit.

{\em Single\/} particle stochastic dynamics conditioned on non-absorption has been explored extensively in probability theory. A regularly updated bibliography is available in Ref.\ \cite{Pollett2012}.
Usually, the conditioning suggests itself in the frame of a biological \cite{Hastings2004}, demographic \cite{Steinsaltz2004} or epidemiological \cite{Nasell1995} context, where the absorbed diffusion process models the
populations undergoing extinction.
By the conditioning on non-absorption the focus is shifted on the behavior of the long-surviving paths of the process.
It may happen that thus conditioned process converges towards a time-invariant distribution, the so called {\em quasi-stationary distribution\/}. The study of quasi-stationary distributions began with the
seminal work of Yaglom on sub-critical Galton-Watson processes \cite{Yaglom1947}. For various stochastic processes, the results on existence, uniqueness and other properties of quasi-stationary distributions
are reviewed in \cite{Cattiaux2009}. Examples of solvable quasi-stationary distributions are the Brownian motion with constant drift absorbed at the origin \cite{Mandl1961, Ferrari1997}, the absorbed logistic Feller diffusion \cite{Meleard2012}, and the Wright-Fisher diffusion \cite{Huillet2007}. In the demographic context, one specific consequence of the conditioning on non-extinction is the deceleration of the instantaneous mortality rates (mortality rate plateaus) \cite{Steinsaltz2004}.

{\em Many-particle\/} diffusion in one-dimensional channels where the particles are not able to pass each other (\emph{the single-file diffusion}, SFD) occurs in many systems such as narrow biological pores \cite{Hodgkin}, the channel systems of zeolites \cite{HahnKarger, KargerBOOK}, or during the sliding of proteins along the DNA \cite{DNAnature, Hippel, Bressloff}. In these systems, the diffusion of the tracer is slowed down due to the interparticle interactions. The mean-square displacement of the tracer increases with time as $t^{1/2}$ in contrast to its linear increase for the free particle. This was first proved by Harris \cite{Harris}. Since then, the single-file diffusion was analytically investigated in many different settings including systems in thermodynamic limit \cite{Jepsen, Percus, Levitt}, infinite line with a finite number of particles \cite{KargerHahn, Kumar, Aslangul}, finite interval \cite{SFDLizana, SFDLizanaPRE, Ryabov2013}, particles under the action of external force field \cite{Silbey, RC2011}. The first-passage problem for a tracer in an infinite system was studied in \cite{SandersAmbjornsson}. The present paper addresses a different setting: in Ref.\ \cite{SandersAmbjornsson} only the tracer feels the absorbing boundary whereas, in the present paper, each particle
can be absorbed. Numerically, the first-passage problem in a driven SFD system was studied in \cite{Barma}.

The present paper continues the study of the model introduced in Ref.\ \cite{RC2012}, where we have investigated the single-file diffusion model including an absorbing boundary. In that paper, we were interested in
the time of absorption of the individual particles. In the present paper the central issue is the tracer dynamics.
The two prerequisites, i.e., the hard-core interaction and the absorption, are essential for a proper understanding of kinetics of diffusion-limited chemical reactions in crowded environments
\cite{DaJiangLiu, Smoluch, JoonhoPark, Kazuhiko}.

The paper is organized as follows. Sec.\ \ref{sec:model} contains the definition of the model. In Sec.\ \ref{sec:single particle}, in order to keep the paper self-contained, we present the complete solution of the underlying single-particle case. Secs.\ \ref{sec:interacting particles}, \ref{sec:TDlimit} comprise our main results.
We first study the unconditioned dynamics of the tracer (Subsecs.\ \ref{subsec:unconditioned dynamics}, \ref{VA}) and then we condition this dynamics on non-absorption (Subsecs.\ \ref{subsec:conditioned dynamics}, \ref{VB}).
\section{\label{sec:model}Definition of the model}
\begin{figure}[t]
\includegraphics[scale=0.8]{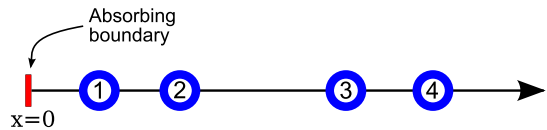}
\caption{\label{fig:fig1}(Color online) Schematic illustration of the possible initial positions of $N=4$ particles and their labeling.}
\end{figure}

Consider the diffusion of hard-core interacting Brownian particles in a semi-infinite one-dimensional interval with the absorbing boundary at the origin. Initially, $N$ particles are distributed along the half-line $(0,+\infty)$. During the time evolution, each particle diffuses with the same diffusion constant, $D$. The particles cannot enter the interval from the outside and they are allowed to leave it only through the boundary at the origin.
The boundary is perfectly absorbing, i.e., if any particle hits the origin it is absorbed with the probability one. At the initial time $t\! =\! 0$, let us label the particles according to ordering of their positions from the left to the right (cf.\ Fig.\ \ref{fig:fig1}). We have
\begin{equation}
\label{ordering}
 0 < \mathbf{X}_{1}(0)<\mathbf{X}_{2}(0)< \ldots <\mathbf{X}_{N}(0) < +\infty \,\,,
\end{equation}
where the random variable $\mathbf{X}_{n}(0)$ denotes the position of the $n$-th particle at $t=0$. The hard-core interaction guarantees that the initial ordering of particles is conserved over time.
Notice that the particle No. $1$ is the first one that might be absorbed. It is only after this event that the particle No. $2$ can approach the origin and be absorbed. Let us denote as $\mathbf{T}_{n}$ the (random) time of the absorption
of the $n$-th particle. Then we have
\begin{equation}
\label{tordering}
 0 < \mathbf{T}_{1}<\mathbf{T}_{2}< \ldots <\mathbf{T}_{N} < +\infty \,\,.
\end{equation}
The last inequality, $\mathbf{T}_{N} < +\infty$, means that the rightmost particle (and hence any particle) is eventually absorbed with the probability one \cite{RC2012}. At the same time, the mean value $\left< \mathbf{T}_{N} \right>$ is infinite \cite{RC2012}.
\section{\label{sec:single particle}Single diffusing particle}
\subsection{Brownian motion absorbed at the origin}

Let us take $N\!=\!1$. Suppose that at the initial time $t=0$ the particle is located at the position $y$, $y>0$. The PDF of the particle's position at the time $t$ conditioned on its initial position is determined by solving the diffusion equation
\begin{equation}
\label{diffeq}
\frac{\partial}{\partial t}f(x;t\,|\,y;0) = D \frac{\partial^2}{\partial x^2}f(x;t\,|\,y;0)\,\,
\end{equation}
subject to the absorbing boundary condition at the origin,
\begin{equation}
\label{absBC}
f(0;t\,|\,y;0) = 0\,\,,
\end{equation}
and to the initial condition
\begin{equation}
f(x;0\,|\,y;0) = \delta(x-y)\,\,.
\end{equation}
This problem is readily solved by the method of images \cite{Chandrasekhar}. The result reads
\begin{equation}
\label{GF}
f(x;t\,|\,y;0)\! =\! \frac{1}{\sqrt{4\pi Dt}}\! \left[\, \displaystyle
{\rm e}^{  -{\left(x-y\right)^{2}}/{4Dt}} -
{\rm e}^{  -{\left(x+y\right)^{2}}/{4Dt}}
\right] \,\,.
\end{equation}
Having this Green's function, the time evolution of an arbitrary initial PDF, say $f(x;0)$, is given by
\begin{equation}
\label{fgenini}
f(x;t)=\left<f(x;t\,|\,\mathbf{X}(0);0)\right>=\int_{0}^{+\infty}\!\!\!\!\!\!\! {\rm d} y\, f(x;t\,|\,y;0) f(y;0)\,\,.
\end{equation}
As for the initial PDF $f(x;0)$, we only assume that all its moments exist. For the sake of illustrations, we take the particular initial condition
\begin{equation}
\label{inicond}
f(x;0) = \frac{ {\rm e}^{-x/L}  }{L}\,\,, \quad L>0\,\,.
\end{equation}

The spatial integral of PDF\ (\ref{fgenini}) over the interval $(0,+\infty)$ equals the \emph{survival probability}, that is, the probability that the particle has not been absorbed by  the time $t$.
If we denote by $\mathbf{T}$ the time of the absorption, we can write
\begin{equation}
S(t)={\rm Prob}\left\{\mathbf{T}>t\right\}=\! \int_{0}^{+\infty} \!\!\!\!\!\!\! {\rm d} x\, f(x;t)=\left<\! {\rm erf}\!  \left( \frac{\mathbf{X}(0)}{\sqrt{4Dt}} \right)\! \right>,
\label{Sgenini}
\end{equation}
where the last expression stands for the average of the error function \cite{AbrStegun} with respect to the initial PDF $f(x;0)$. At the time $t\!=\!0$, the survival probability equals to one.
The long-time behavior of $S(t)$ may be derived by inserting the power series representation of the the error function into Eq.\ (\ref{Sgenini}). The expansion is given in Eq.\ (\ref{Sasy}) and its first term yields
the power-law decay
\begin{equation}
\label{Sdecay}
S(t)\sim\frac{\left< \mathbf{X}(0)\right>}{\sqrt{\pi D }}\, t^{-1/2} \,\,,\qquad t\to\infty\,\,.
\end{equation}
The prefactor depends on the diffusion constant and on the average initial position of the particle. The sign ``$\sim $'' means ``is asymptotically equal''.
\subsection{Brownian motion conditioned on non-absorption}

According to Eq.\ (\ref{Sdecay}) the particle will ultimately hit the absorbing boundary at the origin with probability one. Let us now focus on the dynamics of the particle conditioned on non-absorption.
By definition
\begin{equation}
\label{fconddef}
 f(x;Dt\,|\,\mathbf{T}>t )\! =\! \frac{ f(x;t)}{S(t)} \,\,,
\end{equation}
is the PDF for the particle's position at the time $t$ under the condition that the particle has not been absorbed by the time $t$. The power series representation of PDF $f(x;Dt\,|\,\mathbf{T}>t)$ is given in Eq.\ (\ref{survdenslong}). It follows that
\begin{equation}
\label{survdensshort}
 f(x;Dt\,|\,\mathbf{T}>t)\! = \frac{x}{2Dt}\,
{\rm e}^{ - x^{2}/4Dt} \!
\left(1+\! \textit{O} \left( t^{-1} \right) \right)\,\,,
\end{equation}
where ``$\textit{O} \left( t^{-1} \right)$'' stands for all terms of the series (\ref{survdenslong}) that tend to zero at least as $t^{-1}$ when $t\rightarrow \infty$.
Therefore, in the long-time limit, the PDF (\ref{fconddef}) can be represented by
\begin{equation}
 f^{\rm ( as)}(x;Dt\,|\,\mathbf{T}>t)  =
\frac{x}{2Dt}\,
{\rm e}^{ - x^{2}/4Dt}\,\,.
\label{asyf}
\end{equation}
Notice that this asymptotic representation is non-negative and it is normalized to one on $x\in (0,+\infty)$. The distribution with PDF (\ref{asyf}) is known as the Rayleigh distribution \cite{HandbookOfDistr}.
Further, the asymptotic PDF (\ref{asyf}) is independent of the initial condition $f(x;0)$ and there remains just one length scale associated with the dynamics, the diffusion length $\sqrt{2Dt}$.
All other length scales which have been introduced by the initial condition are already forgotten.

The first and the second moment of the asymptotic PDF (\ref{asyf}) read
\begin{eqnarray}
\label{asymean}
\left< \mathbf{X}(t) \right>^{\rm (as)}_{ \mathbf{T}>t} &=& \sqrt{\pi Dt}\,\,,\\
\left< \mathbf{X}^{2}(t) \right>^{\rm (as)}_{ \mathbf{T}>t} &=& 4 Dt\,\,.
\end{eqnarray}
The mean position of the surviving trajectories should be compared with the corresponding result for the unconditioned dynamics, that is, with $\left< \mathbf{X}(t) \right>=\left< \mathbf{X}(0) \right>$.
Provided a trajectory has avoided absorption by the time $t$, it is typically found far from the origin and its typical position growths as $t^{1/2}$.
The first-order correction to asymptotic result (\ref{asymean}) vanishes as $t^{-1/2}$ and it depends on the initial condition. Using again Eq.\ (\ref{survdenslong}), we get
\begin{equation}
\label{asymeancorrection}
\left< \mathbf{X}(t) \right>_{ \mathbf{T}>t}\!\! = \sqrt{\pi D t}
\left(\! 1 +\frac{1}{12} \frac{\left< \mathbf{X}^{3} (0) \right>}
{\left< \mathbf{X}(0) \right>}\frac{1}{Dt} + \textit{O} \left( t^{-2} \right) \! \right) \,\,.
\end{equation}
\section{\label{sec:interacting particles}$N$ interacting particles}
\subsection{\label{subsec:unconditioned dynamics}Tracer dynamics with absorption}

Considering a general number of particles, $N$, the model setting must be completed by the specification of an initial state. We assume the initial joint probability density function for the positions of the particles vanishes outside the domain $0< x_{1}< ...\, <x_{N}<+\infty$, and, inside this domain, it is given by
\begin{equation}
p(x_{1},x_{2},...\,,x_{N};0) = N! \! \prod_{n=1}^{N} f(x_{n};0) \,\,.
\end{equation}
Throughout the paper, all PDFs that have originated in the single-particle problem are denoted by the letter ``$f$''.
On the other hand, the PDFs in the present $N$-particle problem will be designated by ``$p$''. It is a simple consequence of the assumed hard-core interaction that all $N$-particle quantities can be expressed solely through the single-particle PDFs.

The exact PDF for the position of the $n$-th particle, $n=1,\ldots,N$, reads \footnote{See Eq.\ (35) in Ref.\ \cite{RC2012}}
\begin{eqnarray}
\label{tracerPDF}
p_{\, n}&&(x;t)  = \frac{N!}{(n-1)!(N-n)!}\, f(x;t) \times  \\
&& \times \left( 1-S(t) +\! \int_{0}^{x}\!\!\!\!{\rm d}x' f(x';t) \right)^{n-1}
\!\!\! \left(  \int_{x}^{+\infty}\!\!\!\!\!\!\!\!\!\! {\rm d}x' f(x';t) \right)^{N-n}
\nonumber .
\end{eqnarray}
Apart from the particle labeling, space-time trajectories of hard-core interacting particles are the same as trajectories of noninteracting particles. Hence the probabilistic interpretation behind Eq.\ (\ref{tracerPDF}) can be based on the noninteracting picture. In this picture, the right-hand side (multiplied by ${\rm d}x$) gives the probability of finding a particle in the interval $(x,x+{\rm d}x)$ and, simultaneously, having $(N-n)$ particles to the right of $x$ and $(n-1)$ particles to the left (including those already absorbed by the boundary). The combinatorial factor accounts for all possible labelings of noninteracting particles.

Notice that
\begin{equation}
\label{sumofPDF}
\sum_{n=1}^{N}p_{\, n}(x;t) = N f(x;t)\,\,,
\end{equation}
which can be proved by the direct summation of the expressions (\ref{tracerPDF}). In consequence, this equation tells us that the density of particles for the system of $N$ interacting particles is the same as that for
the system of $N$ noninteracting particles. This holds true for all collective properties. However, the dynamics of the {\em individual\/} particles in the two problems is substantially different.

Let us now derive the long-time asymptotics of the  PDFs (\ref{tracerPDF}). The right-most particle is special. In the long-time limit it behaves in a similar way as the single-diffusing particle \cite{RC2012}.
In particular, for $n=N$, the binomial theorem yields
\begin{eqnarray}
\label{firstparticleasy}
p_{\, N}&& (x;t)  = N f(x;t) \times \\
\nonumber
&& \times \!\! \left[ 1 + (N\! -\! 1)S(t)
\left( \int_{0}^{x}\!\!\!\!{\rm d}x' \frac{f(x';t)}{S(t)}-1 \right)
+ \textit{O}\left(t^{-1}\right) \right] \,,
\end{eqnarray}
where the remaining $(N-2)$ terms of the binomial expansion vanish at least as $(S(t))^{2}$. The integral in (\ref{firstparticleasy}) has been estimated in Eq.\ (\ref{Fconvergence}). On the whole, we obtain
\begin{eqnarray}
\label{firstparticleasy2}
p_{\, N} (x;t) && = N f(x;t) - \\
\nonumber
&&\!\!\! - N (N\! -\! 1)\frac{\left< \mathbf{X}(0) \right>^{2}}{2\pi (Dt)^{2}}
\,x\, {\rm e}^{-x^{2}/2Dt}
+\textit{O}\left(t^{-5/2}\right)\,\,.
\end{eqnarray}
The expression has been written in a way which shows the main asymptotics,
\begin{equation}
p_{\, N}(x; t)\! \sim N  f(x;t) \,\,,
\end{equation}
and the first correction, the second term in (\ref{firstparticleasy2}). The correction describes the relaxation towards the main asymptotics and it is negative.
\begin{figure}[t]
\includegraphics[scale=0.85]{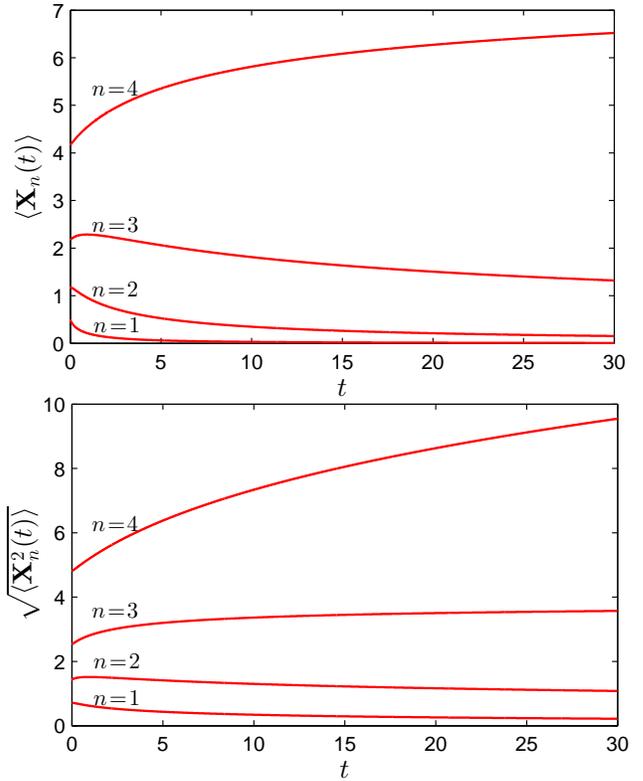}
\caption{\label{fig:fig2}(Color online)
Mean positions (the upper panel) and square roots of the second moments (the lower panel) for the individual particles in the system of $N=4$ interacting particles.
In the underlying single particle problem, we took $D=1$ and the initial condition (\ref{inicond}) with $L=2$.}
\end{figure}

We proceed to the long-time behavior of the $n$-th particle with $n=1,\ldots,(N-1)$. We start again with Eq.\ (\ref{tracerPDF}) and rewrite it in the form
\begin{eqnarray}
\label{pnasy0}
&&p_{\, n}(x;t)=n\binom{N}{n} (S(t))^{N-n}f(x;t) \times  \\
&&\times \left( 1-S(t) +\! \int_{0}^{x}\!\!\!\!{\rm d}x' f(x';t) \right)^{n-1}
\!\!\! \left(1-\!\!\int_{0}^{x}\!\!\!\!\! {\rm d}x'\frac{f(x';t)}{S(t)}\right)^{N-n}\!\!.\nonumber
\end{eqnarray}
The first bracket is again expanded according to the binomial theorem, the second bracket is treated using Eq.\ (\ref{Fconvergence}). Using further Eq.\ (\ref{survdensshort}), the
main asymptotics assumes the form
\begin{equation}
p_{\, n}(x;t)\sim\binom{N}{n}(S(t))^{N-n+1}\,\frac{n x}{2Dt}\,{\rm e}^{-(N-n+1)x^{2}/(4Dt)}\,\,.
\label{pnasy1}
\end{equation}
If we introduce the \emph{renormalized diffusion coefficient},
\begin{equation}
D_{n} = D/(N-n+1)\,\,,
\end{equation}
an important conclusion emerges. On the right-hand side of Eq.\ (\ref{pnasy1}), one recognizes the asymptotic single-particle PDF (\ref{asyf}) conditioned on non-absorption:
\begin{equation}
p_{\, n}(x;t)\! \sim \binom{N}{n-1} (S(t))^{N-n+1} f^{\rm (as)}(x;D_{n}t\,|\, \mathbf{T}>t )\,\,.
\label{pnasy}
\end{equation}
The only difference is that here we have $D_{n}$ instead of $D$ in (\ref{asyf}). The initial order of the particle, $n$, controls, in a decisive way, the main asymptotics. The smaller $n$, the faster is the decay of PDF (for
a given $x$). As a consequence, in the long-time limit, the sum (\ref{sumofPDF}) is dominated by the PDF $p_{\, N}(x; t)$.

Our next goal is analysis of the mean positions of the individual particles. For the rightmost particle, the calculation is based on Eq.\ (\ref{firstparticleasy2}). We obtain
\begin{equation}
\label{firstparticleposition}
\left< \mathbf{X}_{N}(t) \right> = N \left< \mathbf{X}(t) \right>
- N(N\! -\! 1) \frac{\left< \mathbf{X}(0) \right>^{2}}{\sqrt{8\pi Dt}} +\textit{O}\left(t^{-1}\right) \,\,.
\end{equation}
The main asymptotics is covered by the first term on the right hand side, that is, apart from the multiplication by $N$, the main asymptotics coincides with that for the single particle
where we have $\left< \mathbf{X}(t) \right>=\left< \mathbf{X}(0) \right>$. The second term describes corrections. As for the remaining particles, the interaction changes the mean-position dynamics.
The evaluation of the first moments of the densities (\ref{pnasy}) yields the main asymptotics
\begin{equation}
\label{xasy}
\left< \mathbf{X}_{n}(t) \right> \sim  B_{n}\, t^{-(N-n)/2}\,\,, \quad n=1, \ldots , N,
\end{equation}
with the prefactor
\begin{equation}
B_{n}=(N-n+1)^{-1/2} \binom{N}{n-1} \frac{\left< \mathbf{X}(0) \right>^{N-n+1}}{\left(\pi D \right)^{(N-n)/2}} .
\end{equation}
Thus the initial condition and the total number of particles enters the asymptotics only through the prefactor. Notice that the asymptotics for the $n$-th particle for $n < N$, differs from that for the rightmost particle
(and therefore also from that for the single particle), its mean position asymptotically approaches zero.

In a similar way, we readily obtain the second moments. The results are
\begin{eqnarray}
\label{x1asy}
\left< \mathbf{X}_{N}^{2}(t) \right> &\sim &  N \left< \mathbf{X}^{2}(t) \right> - C_{N-1}\,, \\[2pt]
\label{xnasy}
\left< \mathbf{X}_{n}^{2} (t) \right> &\sim & C_{n}\, t^{-(N-n-1)/2},\quad n=1, \ldots , N-1,
\end{eqnarray}
with the prefactors
\begin{equation}
C_{n}= \frac{4D}{N-n+1}  \binom{N}{n-1} \left( \frac{\left< \mathbf{X}(0) \right>}{\sqrt{\pi D}}\right)^{N-n+1} .
\end{equation}
For the rightmost particle, the main asymptotics is proportional to $\left< \mathbf{X}^{2}(t) \right> \sim 4Dt $. Interestingly, for $n=(N-1)$, the second moment approaches the nonzero value $C_{N-1}$ whereas,
for $n<(N-1)$, the second moment decreases towards zero.

The first and the second moments for the individual particles are illustrated in Fig.\ \ref{fig:fig2}. After multiplying Eq.\ (\ref{sumofPDF}) by $x^{k}$ and integrating, we get the relationship
\begin{equation}
\left< \mathbf{X}^{k}(t) \right> = \frac{1}{N} \sum_{n=1}^{N} \left< \mathbf{X}_{n}^{k}(t) \right>
\,\,,\quad k = 0, 1, 2,\ldots\,\,
\label{SumOfMeans}
\end{equation}
valid for any time. In the asymptotic domain, the main asymptotic of the left hand side is covered by the $n=N$ term on the right hand side. Differently speaking, the main asymptotics of the remaining terms in the sum is subdominant with respect to the main asymptotics of the $n=N$ term.
\subsection{\label{subsec:conditioned dynamics}Tracer dynamics conditioned on non-absorption}
\begin{figure}[t]
\includegraphics[scale=0.54]{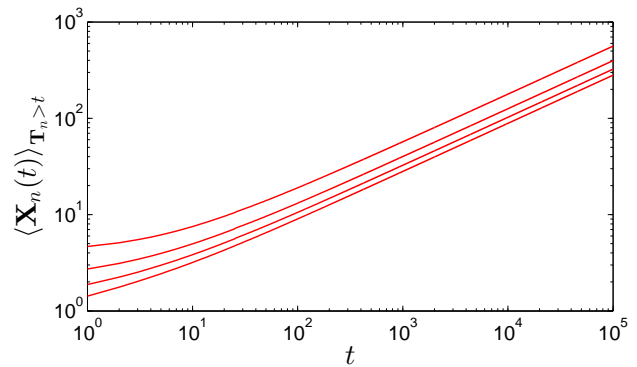}
\caption{\label{fig:fig3}(Color online) The double-logarithmic plot of the conditioned mean positions $\left< \mathbf{X}_{n}(t) \right>_{\mathbf{T}_{n}>t}$ of the individual particles in the system of $N\!=\!4$
interacting particles. The curves are obtained by the numerical integration using PDFs (\ref{conditionedPDF}) with the parameters $L=2$ and $D=1$.
The curves never cross, i.e., the inequalities $\left< \mathbf{X}_{1}(t) \right>_{\mathbf{T}_{1}>t}\!\!\! < \ldots\, <\left< \mathbf{X}_{4}(t) \right>_{\mathbf{T}_{4}>t}$ hold for all $t\geq 0$. The long-time asymptotics does not depend on the initial conditions and it is given by Eq.\ (\ref{survivalmean}).}
\end{figure}

If $\mathbf{T}_{n}$ denotes the time of absorption of the $n$-th particle, its survival probability at the time $t$ is defined by
\begin{equation}
\label{Sndefinition}
S_{n}(t) = {\rm Prob}\left\{\mathbf{T}_{n}>t\right\} = \int_{0}^{+\infty}\!\!\!\!\!\!\! {\rm d}x\, p_{\, n}(x;t)\,\,.
\end{equation}
In \cite{RC2012} we have shown that the above integral can be expressed solely through the survival probability $S(t)$ of the single-diffusing particle, i.e., through the expression (\ref{Sgenini}).
The leading term in the long-time limit is
\begin{equation}
S_{n}(t) =  \binom{N}{n-1} (S(t))^{N-n+1} + \textit{O}\left( t^{-(N-n+2)/2} \right)\,\,.
\label{Snasy}
\end{equation}

Being interested in the long-time dynamics of the individual surviving particles, we introduce the conditional PDFs
\begin{equation}
\label{conditionedPDF}
p_{\, n}(x;Dt\,|\, \mathbf{T}_{n}>t ) = \frac{p_{\, n}(x;t)}{S_{n}(t)} \,\,.
\end{equation}
On the right hand side, the numerator is given in Eq.\ (\ref{tracerPDF}) and the denominator in Eq.\ (\ref{Snasy}). In the long-time limit, the fraction greatly simplifies.
Dividing the main asymptotics (\ref{pnasy}) by the leading term in (\ref{Snasy}) yields
\begin{equation}
\label{conditioneddensitiesasy}
p_{\, n}(x;Dt\,|\,\mathbf{T}_{n}>t )\sim f^{\rm (as)}(x; D_{n}\,t\,|\, \mathbf{T}>t )\,\,.
\end{equation}
This result is remarkable for its simplicity. The asymptotic dynamics of the $n$-th tracer is the same as the dynamics of a single-diffusing particle with the diffusion coefficients $D_{n}=D/(N-n+1)$. In other words, the only implication of the hard-core interaction is the renormalization of the diffusion coefficient.
The left-most particle diffuses with the smallest effective diffusion coefficients $D_{1}=D/N$.
On the other hand the right-most particle has the same effective diffusion coefficient as a single-diffusing particle, $D_{N}=D$.

The above asymptotic relation means that also the moments of the conditioned dynamics are (except of the value of the diffusion coefficient) simply the moments of the single-diffusing particle. More precisely, using Eq.\ (\ref{conditioneddensitiesasy}),
we get
\begin{eqnarray}
\label{survivalmean}
\left<\mathbf{X}_{n}(t) \right>_{\mathbf{T}_{n}>t} &\sim & \sqrt{\pi D_{n} t}\,\,, \\
\label{survivalsecondmoment}
\left<\mathbf{X}_{n}^{2}(t) \right>_{\mathbf{T}_{n}>t} &\sim & 4 D_{n} t\,\,.
\end{eqnarray}
Thus the interparticle interaction {\em does not\/} imply new $n$-dependent dynamical exponents which is the case in the unconditioned dynamics, cf.\ Eqs.\ (\ref{xasy}), (\ref{x1asy}), and (\ref{xnasy}).

Finally, notice that in the present conditioned description there exists no simple relationship similar to Eq.\ (\ref{SumOfMeans}). The $N$-average of the conditioned $k$-th moments is no more equal to the $k$-th conditioned moment for the single-particle diffusion, i.e.,
\begin{equation}
\left< \mathbf{X}^{k}(t)\right>_{\mathbf{T}>t}
\neq
\frac{1}{N} \sum_{n=1}^{N} \left< \mathbf{X}_{n}^{k}(t) \right>_{\mathbf{T}_{n}>t}\,\,.
\end{equation}

\section{\label{sec:TDlimit} Tracer dynamics in thermodynamic limit}
\subsection{\label{VA} Tracer dynamics with absorption}
We now wish to focus on the dynamics of the tracer in a system of infinite number of particles. First, at the initial time $t=0$, the particles are distributed randomly on the half-line with a constant density $\rho$. At the initial instant we activate the absorbing boundary at the origin. Next, the SFD system evolves in time and we are again interested in the dynamics of the $n$-th tagged particle.

Similarly to the previous finite-$N$ case, the analysis is based on the exact PDF for the position of the $n$-th particles. The analytical expression which, in the present context, replaces the formula (\ref{tracerPDF})
reads
\begin{equation}
\label{tracerPDFTD}
p_{n}(x;t) = \frac{\partial \mu(x,t)}{\partial x}\,
\frac{\left[\mu(x,t)\right]^{n-1}}{(n-1)!}\,{\rm e}^{-\mu(x,t)},
\end{equation}
where
\begin{equation}
\label{mudef}
\mu(x,t) = \rho\left[\sqrt{ \frac{4 D t}{\pi}} + \int_{0}^{x} {\rm d}y\, {\rm erf}\!\left(\frac{y}{\sqrt{4Dt}} \right)
\right].
\end{equation}
Notice a straightforward probabilistic interpretation of these formulas. The first term on the right-hand side of Eq.\ (\ref{mudef}), $\rho\sqrt{4 D t/\pi}$, is simply the mean number of the particles absorbed in the time interval $(0,t)$. The second term on the right-hand side represents the mean number of particles which are, at the time $t$, diffusing in the space interval $(0,x)$. Hence, at the time $t$, $\mu(x,t)$ stands for the mean number of particles located to the left of the coordinate $x$, including those which were absorbed. In Eq.\ (\ref{tracerPDFTD}) one recognizes the probability $(\partial \mu/\partial x)\,{\rm d}x$ of finding a noninteracting particle in the interval
$(x,x+{\rm d}x)$ multiplied by the probability that there are $(n-1)$ particles to the left of $x$ (including those already absorbed by the boundary).

The formal derivation of Eq.\ (\ref{tracerPDFTD}) from Eq.\ (\ref{tracerPDF}) proceeds as follows. At the initial time we assume that $N$ particles are homogenously distributed within a finite spatial interval $(0,L)$.
For a large $L$, the probability of finding a single particle to the right of $x$ behaves as
\begin{equation}
\int_{x}^{\infty}\!\!\!\! {\rm d}x' \!\!
\int_{0}^{L}\frac{{\rm d}y}{L}
f(x';t|y;0)
\sim 1 - \frac{1}{L}\, \frac{\mu(x,t)}{\rho}, \quad L\to \infty,
\end{equation}
where $\rho=N/L$. We insert this estimation into Eq.\ (\ref{tracerPDF}). The final result (\ref{tracerPDFTD}) follows after performing the thermodynamic limit: $L\to\infty$, $N\to\infty$, $\rho$ fixed. Interestingly enough,
the passage from Eq.\ (\ref{tracerPDF}) to Eq.\ (\ref{tracerPDFTD}) is similar in spirit to the well known passage from binomial to the Poisson distribution (the law of rare events).

We are again primarily interested in the long-time dynamics of the tracer. After employing an expansion of the integral of the error function in (\ref{mudef}), the main asymptotics of PDF (\ref{tracerPDFTD}) reads
\begin{eqnarray}
\nonumber
p_{n}(x;t) \sim \frac{\left( \rho \sqrt{ \frac{4 D t}{\pi}} \right)^{n-1}}{(n-1)!} \,  {\rm e}^{-\rho \sqrt{4 D t/\pi} }\times \\
\times
\frac{\rho\, x}{\sqrt{\pi D t}}\,
 {\rm e}^{-\rho x^{2}/\sqrt{4 \pi D t}}.
\label{pnTD}
\end{eqnarray}
As for the first two moments of the tracer position we get the asymptotics formulas
\begin{eqnarray}
\label{xnasyTD}
\left< \mathbf{X}_{n}(t) \right> & \sim &\frac{\left( \rho \sqrt{ \frac{4 D t}{\pi}} \right)^{n-1}}{(n-1)!} \,
\sqrt{\frac{\sqrt{\pi^{3} D t}}{2\rho}}
\,{\rm e}^{-\rho \sqrt{4D t/\pi} },\\
\label{xn2asyTD}
\left< \mathbf{X}_{n}^{2}(t) \right> & \sim &
\frac{\left( \rho \sqrt{ \frac{4 D t}{\pi}} \right)^{n-1}}{(n-1)!} \,  \sqrt{\frac{4 \pi D t}{\rho^{2}}}\,{\rm e}^{-\rho \sqrt{4 D t/\pi} }.
\end{eqnarray}
Contrary to the finite-$N$ case (cf.\ Eqs.\ (\ref{firstparticleposition}), (\ref{xasy}), (\ref{x1asy}), and (\ref{xnasy})), the moments vanish for any $n$. The decrease is controlled by a stretched exponential.
\subsection{\label{VB} Tracer dynamics conditioned on non-absorption}
In the thermodynamic limit, the survival probability of the $n$-th tagged particle still depends on its order and, asymptotically, it assumes the form
\begin{equation}
\label{SnTD}
S_{n}(t) \sim \frac{\left( \rho \sqrt{ \frac{4 D t}{\pi}} \right)^{n-1}}{(n-1)!} \,  {\rm e}^{-\rho \sqrt{4 D t/\pi} },
\end{equation}
which is derived by the spatial integration of PDF (\ref{tracerPDFTD}).

Let us now focus on the dynamics of the tracer which has survived by the time $t$. In the large-time limit, the trace PDF conditioned on non-absorption is given by the ratio of asymptotic expressions
(\ref{pnTD}) and (\ref{SnTD}):
\begin{equation}
\label{pnTDCOND}
p_{n}(x;t|\mathbf{T}_{n}>t) \sim
\frac{x}{2 D_{\! 1/2}\sqrt{t}} \,
 {\rm e}^{- x^{2}/(4 D_{\!1/2} \sqrt{t})},
\end{equation}
where we have introduced the generalized diffusion coefficient
\begin{equation}
\label{diffusivity}
D_{\! 1/2} = \sqrt{\frac{\pi D}{4 \rho^{2}}}.
\end{equation}
The asymptotics (\ref{pnTDCOND}) should be contrasted against the single-particle PDF (\ref{asyf}), and the tracer PDF (\ref{conditioneddensitiesasy}) for the finite-$N$ case.
The first two moments of PDF (\ref{pnTDCOND}) are
\begin{eqnarray}
\label{xcond}
\left< \mathbf{X}_{n}(t) \right>_{\mathbf{T}_{n}>t} & \sim & \sqrt{\pi D_{\! 1/2} \sqrt{t} }, \\
\label{x2cond}
\left< \mathbf{X}_{n}^{2}(t) \right>_{\mathbf{T}_{n}>t} & \sim & 4 D_{\! 1/2} \sqrt{ t}.
\end{eqnarray}
Thus the average position of the tracer increases as $t^{1/4}$ in contrast to $t^{1/2}$-law as observed for a finite $N$ (cf.\ Eq.\ (\ref{survivalmean})). The second moment growths as $t^{1/2}$ and hence the tracer dynamics is
subdiffusive.

Finally notice that the generalized diffusion coefficient $D_{\! 1/2}$ is different as compared to the that obtained in a system without the absorbing boundary \cite{Levitt, LeibovichBarkai}.
As it was pointed out in Refs.\ \cite{LizanaBarkai, LeibovichBarkai}, the coefficient $D_{\! 1/2}$ is sensitive to the way the system is prepared. In fact our result (\ref{diffusivity}) indicates that
it also depends on boundary conditions used.
\section{\label{sec:conclusion}Concluding remarks}
Returning to the objectives which were outlined in the Introduction, in the long-time limit the following overall picture emerges.
First, due to the hard-core repulsion, the particle which possesses a right-hand neighbor feels the (moving) reflecting barrier to the right.
The barrier restricts its motion, it reflects the right-moving paths and hence increases the number of left-moving paths.
This left-pushing tendency is illustrated by the asymptotic formulae (\ref{xasy}), (\ref{x1asy}), and (\ref{xnasy}).
The mean position and the mean square displacement of the tracer exhibit new dynamical exponents which depend on its initial order.
Of course, the rightmost particle has no right-hand neighbor and hence it behaves differently.
In the transient regime, its left-hand neighbors still exist and the first particle is pushed to the right. In the asymptotic regime, all other particles have already disappeared
and the first one simply undergoes the free diffusion.
Second, the conditioning on non-absorption removes a part of the left-moving trajectories from the unconditioned ensemble. Hence it imposes, effectively, the right-oriented drift.
Surprisingly enough, the conditioning significantly reduces the effect of the hard-core interaction. The co-operative impact of the both tendencies is behind the asymptotic formulas (\ref{survivalmean}), (\ref{survivalsecondmoment}). The conditioned mean position of the tracer growths as $t^{1/2}$ regardless its order. The interparticle repulsion manifests itself only through
the order-dependent tracer diffusion coefficient. The closer the relative particle position to the boundary the less mobile should that particle be in order to survive for the long times.

The above reasoning holds for the system which initially contains a {\em finite\/} number of particles $N$. In the thermodynamic limit (i.e., assuming initially an {\em infinite\/} number of particles randomly distributed along a half-line with a constant density $\rho$), the long-time dynamics of a tracer is rather different. The new features are based on a simple observation that, for any tracer, there are infinite number of particles to the right of it. This implies the $n$-independent exponential damping of the unconditioned moments (\ref{xnasyTD}), (\ref{xn2asyTD}); the initial order $n$ appears only in the pre-exponential factor.
The conditioned dynamics of a tracer is subdiffusive and independent of $n$ (see Eqs.\ (\ref{pnTDCOND}), (\ref{xcond}), and (\ref{x2cond})). This is in parallel to what has been observed in the SFD without an absorbing
boundary. Namely, for a finite $N$, Aslangul \cite{Aslangul} has shown that, in the long-time limit, the tracer diffusion is normal with the effective diffusion coefficient dependent both on $N$ and on $n$. On the other hand,
for an infinite $N$, one observes an anomalous diffusion and no $n$-dependence \cite{Levitt}. The present paper detects the same features in the SFD with an absorbing boundary.

\begin{acknowledgments}
This work was supported by the grant SVV-2014-267-305, and by the Charles University Grant Agency (project No.\ 301311).
\end{acknowledgments}

\appendix*
\section{Asymptotic expansion of the single-particle PDF conditioned on non-absorption}
The main aim of this Appendix is to justify the relations (\ref{survdensshort}) and (\ref{asymeancorrection}) from the main text.

First we insert the explicit expression (\ref{GF}) into the mean value in Eq.\ (\ref{fgenini}). This yields
\begin{equation}
 f(x;t)\! = \! \frac{2\, {\rm e}^{-x^{2}/4Dt} }{\sqrt{4\pi D t}}\,
\!\!
\left< \!  {\rm sinh}\!\! \left(\frac{ x \mathbf{X}(0)}{2Dt} \right)
{\rm e}^{-\mathbf{X}^{2}\!(0)/4Dt}\! \right> \,\,.
\end{equation}
Using the power series representation for the functions inside the averaging brackets, we obtain
\begin{widetext}
\begin{eqnarray}
\left< \!  {\rm sinh}\!\! \left(\frac{ x \mathbf{X}(0)}{2Dt} \right) {\rm e}^{-\mathbf{X}^{2}\!(0)/4Dt}\! \right>
&=& \frac{x}{2Dt} \sum_{k=0}^{\infty} \sum_{l=0}^{\infty}
\frac{(-1)^{l}2^{-2l}}{l! \left(2k+1 \right)!}
\left( \frac{1}{Dt} \right)^{\! k+l}\!\!\!
 \left( \frac{x^{2}}{4Dt} \right)^{\! k}\!
\left< \mathbf{X}^{2(k+l)+1}\!(0) \right>\,\,.
\label{averseries}
\end{eqnarray}
\end{widetext}
The above double sum is treated by the substitution $p = k+l$:
\begin{equation}
f(x;t)\! =\!\! \frac{x\, {\rm e}^{-x^{2}/4Dt}}{2Dt} \frac{\left< \mathbf{X}(0) \right>}{\sqrt{\pi D t}}
\sum_{p=0}^{\infty}
\sum_{k=0}^{p}\! c(k,p)\!\!
\left( \frac{x^{2}}{4Dt} \right)^{\! k}\!\!\!\!
\left( \frac{1}{Dt} \right)^{\! p} \,\,,
\label{fseries}
\end{equation}
where the time-independent coefficients $c(k,p)$ carry all the information concerning the initial condition. Explicitly, they read
\begin{equation}
c(k,p) =
 \frac{(-1)^{p-k}\, 2^{-2p+2k}}{\left(p-k \right)! \left(2k+1 \right)!}
 \frac{\left<  \mathbf{X}^{2p+1}\!(0) \right>}{\left< \mathbf{X}(0) \right>}
 \,\,.
\label{cprefactor}
\end{equation}

We now prepare similar expansion for the survival probability $S(t)$ as defined in Eq.\ (\ref{Sgenini}). Inserting the power series \cite{AbrStegun}
\begin{equation}
{\rm erf}(z)=\frac{2}{\sqrt{\pi}} \sum_{k=0}^{\infty} \frac{(-1)^{k}z^{2k+1}}{k!\left(2k+1\right)}\,\,
\label{erfseries}
\end{equation}
into the averaging in (\ref{Sgenini}), we immediately obtain
\begin{equation}
\label{Sasy}
S(t) = \frac{\left< \mathbf{X}(0) \right>}{\sqrt{\pi D t}}
\sum_{p=0}^{\infty}
\frac{\left( -1 \right)^{p}}{2^{2p}p\, !(2p+1)}
\frac{\left<  \mathbf{X}^{2p+1}\!(0) \right>}{\left< \mathbf{X}(0) \right>} \!
 \left( \frac{1}{Dt} \right)^{p}\,\,.
\end{equation}
Interestingly, the numerical factors $2^{2p}p\, !(2p+1)$ in the denominators of individual terms of (\ref{Sasy}) form a sequence $1,12,160,2688,55296,...\,$ (for $p = 0,1,2,3,4,\ldots$), which is A167558 sequence in Sloane's On-Line Encyclopedia of Integer Sequences \cite{Sloane}. This sequence originally emerged in a completely different situation  without any obvious connection to the expansion of the error function (see also A167546).

One can arrive at an equivalent expansion of the function $S(t)$ by staring with the first equality in Eq.\ (\ref{Sgenini}) from the main text.
The series (\ref{fseries}) is integrated term by term and it assumes the form
\begin{equation}
S(t) = \frac{\left< \mathbf{X}(0) \right>}{\sqrt{\pi D t}}
\sum_{p=0}^{\infty} \sum_{k=0}^{p}k!c(k,p)
\left( \frac{1}{Dt} \right)^{p}\,\,.
\label{Sasy0}
\end{equation}
By term by term comparison of the series (\ref{Sasy}) and (\ref{Sasy0}) one obtains a non-trivial identity
\begin{equation}
\sum_{k=0}^{p} \frac{(-1)^{k}\, 2^{2k}k\,!}{\left(p-k \right)! \left(2k+1 \right)!}
= \frac{1}{p\,!(2p+1)}\,\,.
\end{equation}

Returning to the main goal of the Appendix, we divide the series (\ref{fseries}) by (\ref{Sasy}). Notice that the prefactor $\left< \mathbf{X}(0) \right>/\sqrt{\pi D t}$ appears in both (\ref{fseries}) and (\ref{Sasy}), therefore it cancels. Representing the fraction $1/S(t)$ by the geometrical series, we finally obtain the sought asymptotic expansion
\begin{widetext}
\begin{equation}
\label{survdenslong}
f(x;Dt\,|\, \mathbf{T}>t)=\frac{x\, }{2Dt}\,{\rm e}^{-x^{2}/4Dt}
\left(1+
 \sum_{k=0}^{1}c(k,1)\!\! \left(\frac{x^{2}}{4Dt}\right)^{\! k}\! \!\! \frac{1}{Dt} +
 \textit{O} \left( t^{-2} \right) \right)\!\!
 \left( 1 + \frac{1}{12}\frac{\left< \mathbf{X}^{3}(0) \right>}{\left< \mathbf{X}(0) \right>}\frac{1}{Dt} +
 \textit{O} \left( t^{-2} \right)  \right)\,\,.
\end{equation}
\end{widetext}
The asymptotic expansion of the corresponding distribution function, i.e.,
\begin{equation}
 \int_{0}^{x}\!\!\!\!{\rm d}x' \frac{f(x';t)}{S(t)} = 1 - {\rm e}^{-x^{2}/4Dt}
 \left(1 + \textit{O} \left( t^{-1} \right)  \right) \,\,,
\label{Fconvergence}
\end{equation}
has been employed in steps leading from Eq.\ (\ref{firstparticleasy}) to Eq.\ (\ref{firstparticleasy2}), and from Eq.\ (\ref{pnasy0}) to Eq.\ (\ref{pnasy1}).

\bibliography{paper}

\providecommand{\noopsort}[1]{}\providecommand{\singleletter}[1]{#1}%
\begin{thebibliography}{42}%
\makeatletter
\providecommand \@ifxundefined [1]{%
 \@ifx{#1\undefined}
}%
\providecommand \@ifnum [1]{%
 \ifnum #1\expandafter \@firstoftwo
 \else \expandafter \@secondoftwo
 \fi
}%
\providecommand \@ifx [1]{%
 \ifx #1\expandafter \@firstoftwo
 \else \expandafter \@secondoftwo
 \fi
}%
\providecommand \natexlab [1]{#1}%
\providecommand \enquote  [1]{``#1''}%
\providecommand \bibnamefont  [1]{#1}%
\providecommand \bibfnamefont [1]{#1}%
\providecommand \citenamefont [1]{#1}%
\providecommand \href@noop [0]{\@secondoftwo}%
\providecommand \href [0]{\begingroup \@sanitize@url \@href}%
\providecommand \@href[1]{\@@startlink{#1}\@@href}%
\providecommand \@@href[1]{\endgroup#1\@@endlink}%
\providecommand \@sanitize@url [0]{\catcode `\\12\catcode `\$12\catcode
  `\&12\catcode `\#12\catcode `\^12\catcode `\_12\catcode `\%12\relax}%
\providecommand \@@startlink[1]{}%
\providecommand \@@endlink[0]{}%
\providecommand \url  [0]{\begingroup\@sanitize@url \@url }%
\providecommand \@url [1]{\endgroup\@href {#1}{\urlprefix }}%
\providecommand \urlprefix  [0]{URL }%
\providecommand \Eprint [0]{\href }%
\providecommand \doibase [0]{http://dx.doi.org/}%
\providecommand \selectlanguage [0]{\@gobble}%
\providecommand \bibinfo  [0]{\@secondoftwo}%
\providecommand \bibfield  [0]{\@secondoftwo}%
\providecommand \translation [1]{[#1]}%
\providecommand \BibitemOpen [0]{}%
\providecommand \bibitemStop [0]{}%
\providecommand \bibitemNoStop [0]{.\EOS\space}%
\providecommand \EOS [0]{\spacefactor3000\relax}%
\providecommand \BibitemShut  [1]{\csname bibitem#1\endcsname}%
\let\auto@bib@innerbib\@empty
\bibitem [{\citenamefont {Pollett}(2012)}]{Pollett2012}%
  \BibitemOpen
  \bibfield  {author} {\bibinfo {author} {\bibfnamefont {P.}~\bibnamefont
  {Pollett}},\ }\href@noop {} {\enquote {\bibinfo {title} {Quasi-stationary
  distributions: a bibliography},}\ }\bibinfo {howpublished} {online at
  http://www.maths.uq.edu.au/~pkp/papers/qsds/qsds.pdf} (\bibinfo {year}
  {2012})\BibitemShut {NoStop}%
\bibitem [{\citenamefont {Hastings}(2004)}]{Hastings2004}%
  \BibitemOpen
  \bibfield  {author} {\bibinfo {author} {\bibfnamefont {A.}~\bibnamefont
  {Hastings}},\ }\href@noop {} {\bibfield  {journal} {\bibinfo  {journal}
  {Trends Ecol. Evol.}\ }\textbf {\bibinfo {volume} {19}},\ \bibinfo {pages}
  {39} (\bibinfo {year} {2004})}\BibitemShut {NoStop}%
\bibitem [{\citenamefont {Steinsaltz}\ and\ \citenamefont
  {Evans}(2004)}]{Steinsaltz2004}%
  \BibitemOpen
  \bibfield  {author} {\bibinfo {author} {\bibfnamefont {D.}~\bibnamefont
  {Steinsaltz}}\ and\ \bibinfo {author} {\bibfnamefont {S.~N.}\ \bibnamefont
  {Evans}},\ }\href@noop {} {\bibfield  {journal} {\bibinfo  {journal} {Theor.
  Pop. Biol.}\ }\textbf {\bibinfo {volume} {65}},\ \bibinfo {pages} {319}
  (\bibinfo {year} {2004})}\BibitemShut {NoStop}%
\bibitem [{\citenamefont {N{\aa}sell}(1995)}]{Nasell1995}%
  \BibitemOpen
  \bibfield  {author} {\bibinfo {author} {\bibfnamefont {I.}~\bibnamefont
  {N{\aa}sell}},\ }\href@noop {} {\bibfield  {journal} {\bibinfo  {journal}
  {Adv. Appl. Probab.}\ }\textbf {\bibinfo {volume} {28}},\ \bibinfo {pages}
  {895} (\bibinfo {year} {1995})}\BibitemShut {NoStop}%
\bibitem [{\citenamefont {Yaglom}(1947)}]{Yaglom1947}%
  \BibitemOpen
  \bibfield  {author} {\bibinfo {author} {\bibfnamefont {A.~M.}\ \bibnamefont
  {Yaglom}},\ }\href@noop {} {\bibfield  {journal} {\bibinfo  {journal} {Dokl.
  Acad. Nauk SSSR (in Russian)}\ }\textbf {\bibinfo {volume} {56}},\ \bibinfo
  {pages} {795} (\bibinfo {year} {1947})}\BibitemShut {NoStop}%
\bibitem [{\citenamefont {Cattiaux}\ \emph {et~al.}(2009)\citenamefont
  {Cattiaux}, \citenamefont {Collet}, \citenamefont {Lambert}, \citenamefont
  {Mart{\' i}nez}, \citenamefont {M{\' e}l{\' e}ard},\ and\ \citenamefont
  {Mart{\' i}n}}]{Cattiaux2009}%
  \BibitemOpen
  \bibfield  {author} {\bibinfo {author} {\bibfnamefont {P.}~\bibnamefont
  {Cattiaux}}, \bibinfo {author} {\bibfnamefont {P.}~\bibnamefont {Collet}},
  \bibinfo {author} {\bibfnamefont {A.}~\bibnamefont {Lambert}}, \bibinfo
  {author} {\bibfnamefont {S.}~\bibnamefont {Mart{\' i}nez}}, \bibinfo {author}
  {\bibfnamefont {S.}~\bibnamefont {M{\' e}l{\' e}ard}}, \ and\ \bibinfo
  {author} {\bibfnamefont {J.~S.}\ \bibnamefont {Mart{\' i}n}},\ }\href@noop {}
  {\bibfield  {journal} {\bibinfo  {journal} {Ann. Probab.}\ }\textbf {\bibinfo
  {volume} {37}},\ \bibinfo {pages} {1926} (\bibinfo {year}
  {2009})}\BibitemShut {NoStop}%
\bibitem [{\citenamefont {Mandl}(1961)}]{Mandl1961}%
  \BibitemOpen
  \bibfield  {author} {\bibinfo {author} {\bibfnamefont {P.}~\bibnamefont
  {Mandl}},\ }\href@noop {} {\bibfield  {journal} {\bibinfo  {journal}
  {Czechoslovak Mathematical Journal}\ }\textbf {\bibinfo {volume} {11}},\
  \bibinfo {pages} {558} (\bibinfo {year} {1961})}\BibitemShut {NoStop}%
\bibitem [{\citenamefont {Ferrari}\ \emph {et~al.}(1997)\citenamefont
  {Ferrari}, \citenamefont {Mart{\' i}nez},\ and\ \citenamefont {Mart{\'
  i}n}}]{Ferrari1997}%
  \BibitemOpen
  \bibfield  {author} {\bibinfo {author} {\bibfnamefont {P.~A.}\ \bibnamefont
  {Ferrari}}, \bibinfo {author} {\bibfnamefont {S.}~\bibnamefont {Mart{\'
  i}nez}}, \ and\ \bibinfo {author} {\bibfnamefont {J.~S.}\ \bibnamefont
  {Mart{\' i}n}},\ }\href@noop {} {\bibfield  {journal} {\bibinfo  {journal}
  {J. Stat. Phys.}\ }\textbf {\bibinfo {volume} {86}},\ \bibinfo {pages} {213}
  (\bibinfo {year} {1997})}\BibitemShut {NoStop}%
\bibitem [{\citenamefont {M{\' e}l{\' e}ard}(2012)}]{Meleard2012}%
  \BibitemOpen
  \bibfield  {author} {\bibinfo {author} {\bibfnamefont {S.}~\bibnamefont {M{\'
  e}l{\' e}ard}},\ }\href@noop {} {\bibfield  {journal} {\bibinfo  {journal}
  {Probability Surveys}\ }\textbf {\bibinfo {volume} {9}},\ \bibinfo {pages}
  {340} (\bibinfo {year} {2012})}\BibitemShut {NoStop}%
\bibitem [{\citenamefont {Huillet}(2007)}]{Huillet2007}%
  \BibitemOpen
  \bibfield  {author} {\bibinfo {author} {\bibfnamefont {T.}~\bibnamefont
  {Huillet}},\ }\href@noop {} {\bibfield  {journal} {\bibinfo  {journal} {J.
  Stat. Mech.-Theory E.}\ }\textbf {\bibinfo {volume} {11}},\ \bibinfo {pages}
  {6} (\bibinfo {year} {2007})}\BibitemShut {NoStop}%
\bibitem [{\citenamefont {Hodgkin}\ and\ \citenamefont
  {Keynes}(1955)}]{Hodgkin}%
  \BibitemOpen
  \bibfield  {author} {\bibinfo {author} {\bibfnamefont {A.~L.}\ \bibnamefont
  {Hodgkin}}\ and\ \bibinfo {author} {\bibfnamefont {R.~D.}\ \bibnamefont
  {Keynes}},\ }\href@noop {} {\bibfield  {journal} {\bibinfo  {journal} {J.
  Physiol.}\ }\textbf {\bibinfo {volume} {128}},\ \bibinfo {pages} {61}
  (\bibinfo {year} {1955})}\BibitemShut {NoStop}%
\bibitem [{\citenamefont {Hahn}\ \emph {et~al.}(1996)\citenamefont {Hahn},
  \citenamefont {K{\" a}rger},\ and\ \citenamefont {Kukla}}]{HahnKarger}%
  \BibitemOpen
  \bibfield  {author} {\bibinfo {author} {\bibfnamefont {K.}~\bibnamefont
  {Hahn}}, \bibinfo {author} {\bibfnamefont {J.}~\bibnamefont {K{\" a}rger}}, \
  and\ \bibinfo {author} {\bibfnamefont {V.}~\bibnamefont {Kukla}},\
  }\href@noop {} {\bibfield  {journal} {\bibinfo  {journal} {Phys. Rev. Lett.}\
  }\textbf {\bibinfo {volume} {76}},\ \bibinfo {pages} {2762} (\bibinfo {year}
  {1996})}\BibitemShut {NoStop}%
\bibitem [{\citenamefont {K{\" a}rger}\ and\ \citenamefont
  {Ruthven}(1992)}]{KargerBOOK}%
  \BibitemOpen
  \bibfield  {author} {\bibinfo {author} {\bibfnamefont {J.}~\bibnamefont {K{\"
  a}rger}}\ and\ \bibinfo {author} {\bibfnamefont {M.}~\bibnamefont
  {Ruthven}},\ }\href@noop {} {\emph {\bibinfo {title} {Diffusion in Zeolites
  and in Other Microporous Solids}}}\ (\bibinfo  {publisher} {Wiley},\ \bibinfo
  {address} {New York},\ \bibinfo {year} {1992})\BibitemShut {NoStop}%
\bibitem [{\citenamefont {Li}\ \emph {et~al.}(2009)\citenamefont {Li},
  \citenamefont {Berg},\ and\ \citenamefont {Elf}}]{DNAnature}%
  \BibitemOpen
  \bibfield  {author} {\bibinfo {author} {\bibfnamefont {G.-W.}\ \bibnamefont
  {Li}}, \bibinfo {author} {\bibfnamefont {O.~G.}\ \bibnamefont {Berg}}, \ and\
  \bibinfo {author} {\bibfnamefont {J.}~\bibnamefont {Elf}},\ }\href@noop {}
  {\bibfield  {journal} {\bibinfo  {journal} {Nature Physics}\ }\textbf
  {\bibinfo {volume} {5}},\ \bibinfo {pages} {294} (\bibinfo {year}
  {2009})}\BibitemShut {NoStop}%
\bibitem [{\citenamefont {Berg}\ \emph {et~al.}(1981)\citenamefont {Berg},
  \citenamefont {Winter},\ and\ \citenamefont {Hippel}}]{Hippel}%
  \BibitemOpen
  \bibfield  {author} {\bibinfo {author} {\bibfnamefont {O.~G.}\ \bibnamefont
  {Berg}}, \bibinfo {author} {\bibfnamefont {R.~B.}\ \bibnamefont {Winter}}, \
  and\ \bibinfo {author} {\bibfnamefont {P.~H.~V.}\ \bibnamefont {Hippel}},\
  }\href@noop {} {\bibfield  {journal} {\bibinfo  {journal} {Biochemistry}\
  }\textbf {\bibinfo {volume} {20}},\ \bibinfo {pages} {6929} (\bibinfo {year}
  {1981})}\BibitemShut {NoStop}%
\bibitem [{\citenamefont {Bressloff}\ and\ \citenamefont
  {Newby}(2013)}]{Bressloff}%
  \BibitemOpen
  \bibfield  {author} {\bibinfo {author} {\bibfnamefont {P.~C.}\ \bibnamefont
  {Bressloff}}\ and\ \bibinfo {author} {\bibfnamefont {J.~M.}\ \bibnamefont
  {Newby}},\ }\href@noop {} {\bibfield  {journal} {\bibinfo  {journal} {Rev.
  Mod. Phys.}\ }\textbf {\bibinfo {volume} {85}},\ \bibinfo {pages} {135}
  (\bibinfo {year} {2013})}\BibitemShut {NoStop}%
\bibitem [{\citenamefont {Harris}(1965)}]{Harris}%
  \BibitemOpen
  \bibfield  {author} {\bibinfo {author} {\bibfnamefont {T.~E.}\ \bibnamefont
  {Harris}},\ }\href@noop {} {\bibfield  {journal} {\bibinfo  {journal} {J.
  Appl. Prob.}\ }\textbf {\bibinfo {volume} {2}},\ \bibinfo {pages} {323}
  (\bibinfo {year} {1965})}\BibitemShut {NoStop}%
\bibitem [{\citenamefont {Jepsen}(1965)}]{Jepsen}%
  \BibitemOpen
  \bibfield  {author} {\bibinfo {author} {\bibfnamefont {D.~W.}\ \bibnamefont
  {Jepsen}},\ }\href@noop {} {\bibfield  {journal} {\bibinfo  {journal} {J.
  Math. Phys.}\ }\textbf {\bibinfo {volume} {6}},\ \bibinfo {pages} {405}
  (\bibinfo {year} {1965})}\BibitemShut {NoStop}%
\bibitem [{\citenamefont {Lebowitz}\ and\ \citenamefont
  {Percus}(1967)}]{Percus}%
  \BibitemOpen
  \bibfield  {author} {\bibinfo {author} {\bibfnamefont {J.~L.}\ \bibnamefont
  {Lebowitz}}\ and\ \bibinfo {author} {\bibfnamefont {J.~K.}\ \bibnamefont
  {Percus}},\ }\href@noop {} {\bibfield  {journal} {\bibinfo  {journal} {Phys.
  Rev.}\ }\textbf {\bibinfo {volume} {155}},\ \bibinfo {pages} {122} (\bibinfo
  {year} {1967})}\BibitemShut {NoStop}%
\bibitem [{\citenamefont {Levitt}(1973)}]{Levitt}%
  \BibitemOpen
  \bibfield  {author} {\bibinfo {author} {\bibfnamefont {D.~G.}\ \bibnamefont
  {Levitt}},\ }\href@noop {} {\bibfield  {journal} {\bibinfo  {journal} {Phys.
  Rev. A}\ }\textbf {\bibinfo {volume} {8}},\ \bibinfo {pages} {3050} (\bibinfo
  {year} {1973})}\BibitemShut {NoStop}%
\bibitem [{\citenamefont {R{\" o}denbeck}\ \emph {et~al.}(1998)\citenamefont
  {R{\" o}denbeck}, \citenamefont {K{\" a}rger},\ and\ \citenamefont
  {Hahn}}]{KargerHahn}%
  \BibitemOpen
  \bibfield  {author} {\bibinfo {author} {\bibfnamefont {C.}~\bibnamefont {R{\"
  o}denbeck}}, \bibinfo {author} {\bibfnamefont {J.}~\bibnamefont {K{\"
  a}rger}}, \ and\ \bibinfo {author} {\bibfnamefont {K.}~\bibnamefont {Hahn}},\
  }\href@noop {} {\bibfield  {journal} {\bibinfo  {journal} {Phys. Rev. E}\
  }\textbf {\bibinfo {volume} {57}},\ \bibinfo {pages} {4382} (\bibinfo {year}
  {1998})}\BibitemShut {NoStop}%
\bibitem [{\citenamefont {Kumar}(2008)}]{Kumar}%
  \BibitemOpen
  \bibfield  {author} {\bibinfo {author} {\bibfnamefont {D.}~\bibnamefont
  {Kumar}},\ }\href@noop {} {\bibfield  {journal} {\bibinfo  {journal} {Phys.
  Rev. E}\ }\textbf {\bibinfo {volume} {78}},\ \bibinfo {pages} {021133}
  (\bibinfo {year} {2008})}\BibitemShut {NoStop}%
\bibitem [{\citenamefont {Aslangul}(1998)}]{Aslangul}%
  \BibitemOpen
  \bibfield  {author} {\bibinfo {author} {\bibfnamefont {C.}~\bibnamefont
  {Aslangul}},\ }\href@noop {} {\bibfield  {journal} {\bibinfo  {journal}
  {Europhys. Lett.}\ }\textbf {\bibinfo {volume} {44}},\ \bibinfo {pages} {284}
  (\bibinfo {year} {1998})}\BibitemShut {NoStop}%
\bibitem [{\citenamefont {Lizana}\ and\ \citenamefont {Ambj{\"
  o}rnsson}(2008)}]{SFDLizana}%
  \BibitemOpen
  \bibfield  {author} {\bibinfo {author} {\bibfnamefont {L.}~\bibnamefont
  {Lizana}}\ and\ \bibinfo {author} {\bibfnamefont {T.}~\bibnamefont {Ambj{\"
  o}rnsson}},\ }\href@noop {} {\bibfield  {journal} {\bibinfo  {journal} {Phys.
  Rev. Lett.}\ }\textbf {\bibinfo {volume} {100}},\ \bibinfo {pages} {200601}
  (\bibinfo {year} {2008})}\BibitemShut {NoStop}%
\bibitem [{\citenamefont {Lizana}\ and\ \citenamefont {Ambj{\"
  o}rnsson}(2009)}]{SFDLizanaPRE}%
  \BibitemOpen
  \bibfield  {author} {\bibinfo {author} {\bibfnamefont {L.}~\bibnamefont
  {Lizana}}\ and\ \bibinfo {author} {\bibfnamefont {T.}~\bibnamefont {Ambj{\"
  o}rnsson}},\ }\href@noop {} {\bibfield  {journal} {\bibinfo  {journal} {Phys.
  Rev. E}\ }\textbf {\bibinfo {volume} {80}},\ \bibinfo {pages} {051103}
  (\bibinfo {year} {2009})}\BibitemShut {NoStop}%
\bibitem [{\citenamefont {Ryabov}(2013)}]{Ryabov2013}%
  \BibitemOpen
  \bibfield  {author} {\bibinfo {author} {\bibfnamefont {A.}~\bibnamefont
  {Ryabov}},\ }\href@noop {} {\bibfield  {journal} {\bibinfo  {journal} {J.
  Chem. Phys.}\ }\textbf {\bibinfo {volume} {138}},\ \bibinfo {pages} {154104}
  (\bibinfo {year} {2013})}\BibitemShut {NoStop}%
\bibitem [{\citenamefont {Barkai}\ and\ \citenamefont {Silbey}(2009)}]{Silbey}%
  \BibitemOpen
  \bibfield  {author} {\bibinfo {author} {\bibfnamefont {E.}~\bibnamefont
  {Barkai}}\ and\ \bibinfo {author} {\bibfnamefont {R.}~\bibnamefont
  {Silbey}},\ }\href@noop {} {\bibfield  {journal} {\bibinfo  {journal} {Phys.
  Rev. Lett.}\ }\textbf {\bibinfo {volume} {102}},\ \bibinfo {pages} {050602}
  (\bibinfo {year} {2009})}\BibitemShut {NoStop}%
\bibitem [{\citenamefont {Ryabov}\ and\ \citenamefont
  {Chvosta}(2011)}]{RC2011}%
  \BibitemOpen
  \bibfield  {author} {\bibinfo {author} {\bibfnamefont {A.}~\bibnamefont
  {Ryabov}}\ and\ \bibinfo {author} {\bibfnamefont {P.}~\bibnamefont
  {Chvosta}},\ }\href@noop {} {\bibfield  {journal} {\bibinfo  {journal} {Phys.
  Rev. E.}\ }\textbf {\bibinfo {volume} {83}},\ \bibinfo {pages} {020106}
  (\bibinfo {year} {2011})}\BibitemShut {NoStop}%
\bibitem [{\citenamefont {Sanders}\ and\ \citenamefont {Ambj{\"
  o}rnsson}(2012)}]{SandersAmbjornsson}%
  \BibitemOpen
  \bibfield  {author} {\bibinfo {author} {\bibfnamefont {L.~P.}\ \bibnamefont
  {Sanders}}\ and\ \bibinfo {author} {\bibfnamefont {T.}~\bibnamefont {Ambj{\"
  o}rnsson}},\ }\href@noop {} {\bibfield  {journal} {\bibinfo  {journal} {J.
  Chem. Phys.}\ }\textbf {\bibinfo {volume} {136}},\ \bibinfo {pages} {175103}
  (\bibinfo {year} {2012})}\BibitemShut {NoStop}%
\bibitem [{\citenamefont {Barma}\ and\ \citenamefont
  {Ramaswamy}(1986)}]{Barma}%
  \BibitemOpen
  \bibfield  {author} {\bibinfo {author} {\bibfnamefont {M.}~\bibnamefont
  {Barma}}\ and\ \bibinfo {author} {\bibfnamefont {R.}~\bibnamefont
  {Ramaswamy}},\ }\href@noop {} {\bibfield  {journal} {\bibinfo  {journal} {J.
  Stat. Phys.}\ }\textbf {\bibinfo {volume} {43}},\ \bibinfo {pages} {561}
  (\bibinfo {year} {1986})}\BibitemShut {NoStop}%
\bibitem [{\citenamefont {Ryabov}\ and\ \citenamefont
  {Chvosta}(2012)}]{RC2012}%
  \BibitemOpen
  \bibfield  {author} {\bibinfo {author} {\bibfnamefont {A.}~\bibnamefont
  {Ryabov}}\ and\ \bibinfo {author} {\bibfnamefont {P.}~\bibnamefont
  {Chvosta}},\ }\href@noop {} {\bibfield  {journal} {\bibinfo  {journal} {J.
  Chem. Phys.}\ }\textbf {\bibinfo {volume} {136}},\ \bibinfo {pages} {064114}
  (\bibinfo {year} {2012})}\BibitemShut {NoStop}%
\bibitem [{\citenamefont {Liu}\ \emph {et~al.}(2011)\citenamefont {Liu},
  \citenamefont {Wang}, \citenamefont {Ackerman}, \citenamefont {Slowing},
  \citenamefont {Pruski}, \citenamefont {Chen}, \citenamefont {Lin},\ and\
  \citenamefont {Evans}}]{DaJiangLiu}%
  \BibitemOpen
  \bibfield  {author} {\bibinfo {author} {\bibfnamefont {D.-J.}\ \bibnamefont
  {Liu}}, \bibinfo {author} {\bibfnamefont {J.}~\bibnamefont {Wang}}, \bibinfo
  {author} {\bibfnamefont {D.~M.}\ \bibnamefont {Ackerman}}, \bibinfo {author}
  {\bibfnamefont {I.~I.}\ \bibnamefont {Slowing}}, \bibinfo {author}
  {\bibfnamefont {M.}~\bibnamefont {Pruski}}, \bibinfo {author} {\bibfnamefont
  {H.-T.}\ \bibnamefont {Chen}}, \bibinfo {author} {\bibfnamefont {V.~S.-Y.}\
  \bibnamefont {Lin}}, \ and\ \bibinfo {author} {\bibfnamefont {J.~W.}\
  \bibnamefont {Evans}},\ }\href@noop {} {\bibfield  {journal} {\bibinfo
  {journal} {ACS Catal.}\ }\textbf {\bibinfo {volume} {1}},\ \bibinfo {pages}
  {751} (\bibinfo {year} {2011})}\BibitemShut {NoStop}%
\bibitem [{\citenamefont {Dorsaz}\ \emph {et~al.}(2010)\citenamefont {Dorsaz},
  \citenamefont {Michele}, \citenamefont {Piazza}, \citenamefont {Rios},\ and\
  \citenamefont {Foffi}}]{Smoluch}%
  \BibitemOpen
  \bibfield  {author} {\bibinfo {author} {\bibfnamefont {N.}~\bibnamefont
  {Dorsaz}}, \bibinfo {author} {\bibfnamefont {C.~D.}\ \bibnamefont {Michele}},
  \bibinfo {author} {\bibfnamefont {F.}~\bibnamefont {Piazza}}, \bibinfo
  {author} {\bibfnamefont {P.~D.~L.}\ \bibnamefont {Rios}}, \ and\ \bibinfo
  {author} {\bibfnamefont {G.}~\bibnamefont {Foffi}},\ }\href@noop {}
  {\bibfield  {journal} {\bibinfo  {journal} {Phys. Rev. Lett.}\ }\textbf
  {\bibinfo {volume} {105}},\ \bibinfo {pages} {120601} (\bibinfo {year}
  {2010})}\BibitemShut {NoStop}%
\bibitem [{\citenamefont {Park}\ \emph {et~al.}(2003)\citenamefont {Park},
  \citenamefont {Kim},\ and\ \citenamefont {Shin}}]{JoonhoPark}%
  \BibitemOpen
  \bibfield  {author} {\bibinfo {author} {\bibfnamefont {J.}~\bibnamefont
  {Park}}, \bibinfo {author} {\bibfnamefont {H.}~\bibnamefont {Kim}}, \ and\
  \bibinfo {author} {\bibfnamefont {K.~J.}\ \bibnamefont {Shin}},\ }\href@noop
  {} {\bibfield  {journal} {\bibinfo  {journal} {J. Chem. Phys.}\ }\textbf
  {\bibinfo {volume} {118}},\ \bibinfo {pages} {9697} (\bibinfo {year}
  {2003})}\BibitemShut {NoStop}%
\bibitem [{\citenamefont {Seki}\ \emph {et~al.}(2011)\citenamefont {Seki},
  \citenamefont {Wojcik},\ and\ \citenamefont {Tachiya}}]{Kazuhiko}%
  \BibitemOpen
  \bibfield  {author} {\bibinfo {author} {\bibfnamefont {K.}~\bibnamefont
  {Seki}}, \bibinfo {author} {\bibfnamefont {M.}~\bibnamefont {Wojcik}}, \ and\
  \bibinfo {author} {\bibfnamefont {M.}~\bibnamefont {Tachiya}},\ }\href@noop
  {} {\bibfield  {journal} {\bibinfo  {journal} {J. Chem. Phys.}\ }\textbf
  {\bibinfo {volume} {134}},\ \bibinfo {pages} {094506} (\bibinfo {year}
  {2011})}\BibitemShut {NoStop}%
\bibitem [{\citenamefont {Chandrasekhar}(1943)}]{Chandrasekhar}%
  \BibitemOpen
  \bibfield  {author} {\bibinfo {author} {\bibfnamefont {S.}~\bibnamefont
  {Chandrasekhar}},\ }\href@noop {} {\bibfield  {journal} {\bibinfo  {journal}
  {Rev. Mod. Phys.}\ }\textbf {\bibinfo {volume} {15}},\ \bibinfo {pages} {1}
  (\bibinfo {year} {1943})}\BibitemShut {NoStop}%
\bibitem [{\citenamefont {Abramowitz}\ and\ \citenamefont
  {Stegun}(1972)}]{AbrStegun}%
  \BibitemOpen
  \bibinfo {editor} {\bibfnamefont {M.}~\bibnamefont {Abramowitz}}\ and\
  \bibinfo {editor} {\bibfnamefont {I.~A.}\ \bibnamefont {Stegun}},\ eds.,\
  \href@noop {} {\emph {\bibinfo {title} {Handbook of Mathematical Functions
  with Formulas, Graphs, and Mathematical Tables}}},\ \bibinfo {edition} {9th}\
  ed.\ (\bibinfo  {publisher} {Dover},\ \bibinfo {address} {New York},\
  \bibinfo {year} {1972})\BibitemShut {NoStop}%
\bibitem [{\citenamefont {Krishnamoorthy}(2006)}]{HandbookOfDistr}%
  \BibitemOpen
  \bibfield  {author} {\bibinfo {author} {\bibfnamefont {K.}~\bibnamefont
  {Krishnamoorthy}},\ }\href@noop {} {\emph {\bibinfo {title} {Handbook of
  Statistical Distributions with Applications}}},\ \bibinfo {edition} {1st}\
  ed.\ (\bibinfo  {publisher} {Chapman \& Hall/CRC},\ \bibinfo {address} {New
  York},\ \bibinfo {year} {2006})\BibitemShut {NoStop}%
\bibitem [{Note1()}]{Note1}%
  \BibitemOpen
  \bibinfo {note} {See Eq.\ (35) in Ref.\ \cite {RC2012}}\BibitemShut {NoStop}%
\bibitem [{\citenamefont {Leibovich}\ and\ \citenamefont
  {Barkai}(2013)}]{LeibovichBarkai}%
  \BibitemOpen
  \bibfield  {author} {\bibinfo {author} {\bibfnamefont {N.}~\bibnamefont
  {Leibovich}}\ and\ \bibinfo {author} {\bibfnamefont {E.}~\bibnamefont
  {Barkai}},\ }\href@noop {} {\bibfield  {journal} {\bibinfo  {journal} {Phys.
  Rev. E}\ }\textbf {\bibinfo {volume} {88}},\ \bibinfo {pages} {032107}
  (\bibinfo {year} {2013})}\BibitemShut {NoStop}%
\bibitem [{\citenamefont {Lizana}\ \emph {et~al.}(2010)\citenamefont {Lizana},
  \citenamefont {Ambj{\" o}rnsson}, \citenamefont {Taloni}, \citenamefont
  {Barkai},\ and\ \citenamefont {Lomholt}}]{LizanaBarkai}%
  \BibitemOpen
  \bibfield  {author} {\bibinfo {author} {\bibfnamefont {L.}~\bibnamefont
  {Lizana}}, \bibinfo {author} {\bibfnamefont {T.}~\bibnamefont {Ambj{\"
  o}rnsson}}, \bibinfo {author} {\bibfnamefont {A.}~\bibnamefont {Taloni}},
  \bibinfo {author} {\bibfnamefont {E.}~\bibnamefont {Barkai}}, \ and\ \bibinfo
  {author} {\bibfnamefont {M.~A.}\ \bibnamefont {Lomholt}},\ }\href@noop {}
  {\bibfield  {journal} {\bibinfo  {journal} {Phys. Rev. E}\ }\textbf {\bibinfo
  {volume} {81}},\ \bibinfo {pages} {051118} (\bibinfo {year}
  {2010})}\BibitemShut {NoStop}%
\bibitem [{\citenamefont {Sloane}()}]{Sloane}%
  \BibitemOpen
  \bibfield  {author} {\bibinfo {author} {\bibfnamefont {N.~J.~A.}\
  \bibnamefont {Sloane}},\ }\href@noop {} {\enquote {\bibinfo {title} {The
  on-line encyclopedia of integer sequences},}\ }\bibinfo {howpublished}
  {online at https://oeis.org/}\BibitemShut {NoStop}%
\end{thebibliography}%

\end{document}